\documentclass[twoside]{article}
\usepackage{fleqn,espcrc2}

\newcommand{\AmS}{{\protect\the\textfont2
  A\kern-.1667em\lower.5ex\hbox{M}\kern-.125emS}}

\title{Relic neutrino asymmetry generation from $\nu_{\alpha} \leftrightarrow \nu_s$
oscillations}

\author{Yvonne Y. Y. Wong\address{School of Physics, Research
Centre for High Energy Physics,
        \\
        The University of Melbourne Vic 3010, Australia}%
        \thanks{Talk given at NOW2000.
        Work supported by the Commonwealth of Australia's
        Postgraduate Award Scheme.}}

\begin{document}

\begin{abstract}
Active--sterile neutrino oscillations provide a mechanism by which
large differences in the neutrino and antineutrino number
densities can be created in the early universe. The quantum
kinetic equations
 are employed in the study of these neutrino asymmetries, which,
 when solved analytically in the adiabatic limit, generate
physically transparent evolution equations that are very useful
for the understanding of the  nature of the asymmetry growth.
\vspace{1pc}
\end{abstract}

\maketitle

\section{Introduction}

Active--sterile neutrino oscillations in the early universe can
generate large differences in the number densities of relic
neutrinos and antineutrinos \cite{ftv,longpaper,fv1}. For a wide
range of small vacuum mixing angles $\theta_0$ and squared mass
differences $\Delta m^2$ (provided that $\Delta m^2 < 0$, meaning
roughly that $m_{\nu_{\alpha}}
> m_{\nu_s}$, where $\alpha = e$, $\mu$ or $\tau$),
this is achieved in the temperature range $T \sim 1 \to {\cal O}
(10)$ MeV  (before neutrino decoupling at $T \simeq 1$ MeV) in a
$CP$ asymmetric background.  The final values of these so-called
lepton number asymmetries,
\begin{equation}
L_{\nu_{\alpha}} \equiv \frac{n_{\nu_{\alpha}} -
n_{\overline{\nu}_{\alpha}}}{{n_{\gamma}}},
\end{equation}
where $n_{\psi}$ is the number density of species $\psi$, are
typically of the order $0.1$, and have profound cosmological
implications: (i) the suppression of $\nu_s$ production because of
large matter effects \cite{ftv,longpaper}, and (ii) a modification
to primordial $^4 {\rm He}$ synthesis if a sufficiently large
$L_{\nu_e}$ is created prior to or during big bang nucleosynthesis
(BBN) \cite{fv1}. For a review on asymmetry generation, see Ref.\
\cite{comment}.

The study of asymmetry evolution is conducted in the density
matrix formalism, and the quantum kinetic equations (QKEs)
\cite{qke,qke2}, which quantify the  effects of (i) decohering
collisions, (ii) matter-affected oscillations, (iii) expansion of
the universe, and (iv) repopulation of empty  states from the
background, are the main tool of trade.

\section{Quantum kinetic equations} \label{definitions}

The properties of a $\nu_{\alpha} \leftrightarrow \nu_s$ system
are encoded in the density matrices \cite{qke,qke2}
\begin{equation}
\rho = \frac{1}{2} ( P_0 + {\bf P} \cdot \sigma ),
\end{equation}
where ${\bf P} = P_x \hat{x} + P_y \hat{y} + P_z \hat{z}$. All
quantities are understood to be functions of temperature $T$ and
momentum $p$.  The diagonal entries represent, respectively, the
$\nu_{\alpha}$ and $\nu_s$ distribution functions:
\begin{equation} N_{\nu_{\alpha}} = \frac{1}{2} (P_0 + P_z)N_{{\rm
eq}}^0, \; N_{\nu_s} =\frac{1}{2} (P_0 - P_z) N_{{\rm eq}}^0,
\end{equation}
in which the reference distribution function $N_{{\rm eq}}^0$ is
of Fermi--Dirac (equilibrium) form,
\begin{equation}
N_{{\rm eq}}^{\mu} = \frac{1}{2 \pi^2}
\frac{p^2}{1+\exp\left(\frac{p-\mu}{T}\right)},
\end{equation}
with chemical potential $\mu$ set to zero.  The off-diagonal
entries $P_x$ and $P_y$ are the coherences.

The  variables $P_0$ and ${\bf P}$ advance in time according to
the quantum kinetic equations (QKEs) \cite{qke}
\begin{eqnarray}
\label{qkes} \frac{\partial {\bf P}}{\partial t} &=& {\bf V}
\times {\bf P} - D (P_x \hat{x} + P_y \hat{y}) + R_{\alpha}
\hat{z},\nonumber
\\ \frac{\partial P_0}{\partial t} &=& R_{\alpha} \simeq \Gamma
\left[ \frac{N_{\rm eq}^{\mu}}{N_{{\rm eq}}^0}  - \frac{1}{2}(P_0
+ P_z) \right].
\end{eqnarray}
Here, the matter potential vector ${\bf V} = \beta \hat{x} +
\lambda \hat{z}$ governs the coherent matter-affected evolution of
the ensemble, with \cite{msw,rn}
\begin{equation}
\label{betalambda} \beta  = \frac{\Delta m^2}{2p} \sin 2 \theta_0,
\;  \lambda  = \frac{\Delta m^2}{2p}(b - a - \cos 2 \theta_0),
\end{equation}
and
\begin{eqnarray}
\label{ap} a &=& -\frac{4 \zeta(3) \sqrt{2} G_F L^{(\alpha)} T^3
p}{\pi^2 \Delta m^2}, \nonumber \\ b &=& -\frac{4 \zeta(3)
\sqrt{2} G_F A_{\alpha} T^4 p^2}{\pi^2 \Delta m^2 m^2_W},
\end{eqnarray}
where $G_F$ is the Fermi constant, $m_W$ the $W$-boson mass,
$\zeta$ the Riemann zeta function and $A_e \simeq 17$,
$A_{\mu,\\tau} \simeq 4.9$.  Also present in Eq.\ (\ref{ap}) is
\begin{equation}
\label{efflep} L^{(\alpha)} = L_{\nu_{\alpha}}+L_{\nu_e} +
L_{\nu_{\mu}} + L_{\nu_{\tau}} + \eta,
\end{equation}
which contains a small term due to the cosmological
baryon--antibaryon asymmetry $\eta$.

The decoherence function $D$ is related to the  total collision
rate for $\nu_{\alpha}$, $\Gamma$, via
 $D=\Gamma/2 =
 k_{\alpha} G^2_F T^4 p$, where $k_e \simeq
0.635$, $k_{\mu, \tau} \simeq 0.46$ \cite{qke}, while the
repopulation function $R_{\alpha}$ determines the rate at which a
depleted momentum state in the distribution is refilled from the
background plasma so as to restore thermal equilibrium.

A separate  set of expressions, denoted with an overhead bar,
parameterises the $\overline{\nu}_{\alpha} \leftrightarrow
\overline{\nu}_s$ system.

\section{Adiabatic limit approximation}
\label{adiabatic}

By demanding $\alpha + s$ lepton number conservation,  one may
extract from the QKEs an exact evolution equation for the
asymmetry, that is \cite{bvw},
\begin{equation}
\label{exactdldt} \frac{dL_{\nu_{\alpha}}}{dt} =
\frac{1}{2n_{\gamma}} \int \beta (P_y - \overline{P}_y) N_{{\rm
eq}}^0 dp.
\end{equation}
The role of the adiabatic limit approximation \cite{bvw,vw,lvw} is
to supply approximate analytical solutions for $P_y$ and
$\overline{P}_y$ so as to render Eq.\ (\ref{exactdldt}) a more
physically transparent expression, through which the  nature of
the asymmetry growth may be revealed.

\subsection{The Boltzmann limit}
\label{boltzmannlimit}

The first assumption is a small  refilling function
\cite{bvw,lvw}, i.e., $R_{\alpha} \simeq 0$, so that the QKEs
reduce to
\begin{equation}
\label{pqke} \frac{\partial {\bf P}}{\partial t}  \simeq {\cal
K}{\bf P}
 = \left(
\begin{array}{ccc}
            -D & -\lambda & 0 \\
            \lambda & -D & -\beta \\
            0 & \beta & 0 \end{array} \right)
            \left[ \begin{array}{c}
                    P_x \\
                    P_y \\
                    P_z \end{array} \right].
\end{equation}
This may be solved by establishing an ``instantaneous diagonal
basis'' via ${\bf Q} = {\cal U} {\bf P}$ and ${\cal K}_d \equiv
{\rm diag}(k_1,\, k_2,\, k_3) = {\cal U K U}^{-1}$, with
eigenvalues
\begin{equation}
\label{k1k2} k_{1,2} =  -d \pm i \omega, \quad k_3 =
-\frac{\beta^2 D}{d^2 + \omega^2},
\end{equation}
where
\begin{equation}
\label{dandomega}
 d =
D+\frac{k_3}{2}, \quad \omega^2 = \lambda^2 + \beta^2 + k_3 D +
\frac{3}{4} k_3^2,
\end{equation}
are interpreted as the {\it effective} damping factor and
oscillation frequency respectively \cite{vw}.

In the new basis, Eq.\ (\ref{pqke}) is similar to
\begin{equation}
\label{qDE} \frac{\partial {\bf Q}}{\partial t} = {\cal K}_d {\bf
Q} - {\cal U} \frac{\partial {\cal U}}{\partial t}^{-1} {\bf Q},
\end{equation}
where the second term is explicitly dependent on the time
derivatives of $D$, $\lambda$ and $\beta$. The adiabatic limit is
defined by  ${\cal U} \frac{\partial {\cal U}}{\partial t}^{-1}
\simeq 0$ \cite{bvw}, such that
\begin{equation}
\label{psolution} P_{\delta}(t) \simeq \sum_{i,\epsilon} {\cal
U}^{-1}_{\delta i} (t) e^{\int^t_0 k_i(t') dt'} {\cal U}_{i
\epsilon} (0) P_{\epsilon} (0),
\end{equation}
where $\delta, \epsilon = x,y,z$, and $i=1,2,3$, provides an
approximate solution to Eq.\ (\ref{pqke}).

Equation (\ref{psolution}) simplifies further in the high
temperature collision-dominated $D \gg |\beta|$ limit with the
approximation
\begin{equation}
\label{exp} e^{\int^t_0 k_{1,\,2}(t') dt'} \to 0,
\end{equation}
since, in this limit, $d \simeq D$, $\omega \simeq \lambda$, and
thus $|k_3| \ll |{\rm Re}(k_{1,2})|$ by Eqs.\ (\ref{k1k2}) and
(\ref{dandomega}). Physically, frequent collisions lead to rapid
and complete damping of the system's innate oscillations. The
evolution of the ensemble is dictated instead by a slower rate
$k_3$, which tends to equalise the $\nu_{\alpha}$ and $\nu_s$
distribution functions \cite{vw}.

Specifically, one has, from Eqs.\ (\ref{psolution}) and
(\ref{exp}),
\begin{equation} \label{pypzapprox} P_y(t) \simeq
\frac{{\cal U}_{y3}^{-1}(t)}{{\cal U}_{z3}^{-1}(t)}P_z(t) =
\frac{k_3}{\beta} P_z(t),
\end{equation}
which
allows Eq.\
(\ref{exactdldt}) to be approximated as
\begin{eqnarray}
\label{approxdldt} \frac{dL_{\nu_{\alpha}}}{dt} \simeq \frac{1}{2
n_{\gamma}} && \hspace{-5mm} \int \left[ k_3
(N_{\nu_{\alpha}}-N_{\nu_s}) \right. \nonumber
\\ && \hspace{5mm}- \left. \overline{k}_3
(N_{\overline{\nu}_{\alpha}}-N_{\overline{\nu}_s}) \right] dp.
\end{eqnarray}
Equation (\ref{approxdldt}) is essentially a classical Boltzmann
rate equation that is dependent {\it only} on the instantaneous
$\nu_{\alpha}$ and $\nu_s$ distribution functions. The ``rate
constant'' $k_3$ is of  particular interest:  A {\it matter- and
collision-affected mixing angle} arises naturally from the QKEs
\cite{vw}:
\begin{equation}
\label{angle}
\sin^2 2 \theta_{m,D} \equiv \frac{\beta^2}{(D+
k_3)^2 + \lambda^2 + \beta^2},
\end{equation}
such that $k_3 = - \sin^2 2 \theta_{m,D} \times D$; $k_3$
therefore reflects on the system's ability to mix
 and to collide.

After further algebraic manipulations, Eq.\ (\ref{approxdldt})
shows that, for $\Delta m^2 < 0$, there exists a critical
temperature above which $\frac{dL_{\nu_{\alpha}}}{dt} \propto -
L^{(\alpha)}$, and below which $\frac{dL_{\nu_{\alpha}}}{dt}
\propto + L^{(\alpha)}$. The former implies $L_{\nu_{\alpha}}$
destruction, while the latter causes exponential growth
\cite{longpaper}. These features have been confirmed by numerical
integration of the exact QKEs.

\subsection{Collision-affected MSW effect}

Shortly after the initial explosive growth, the system enters a
low temperature regime in which coherent
Mikheyev--Smirnov--Wolfenstein (MSW) transitions \cite{msw} are
the dominant asymmetry amplifier. Subsequent growth follows an
approximate power law rate $L_{\nu_{\alpha}} \propto T^{-4}$,
reaching a value of ${\cal O} (0.1)$ prior to neutrino decoupling
 \cite{fv1}. In the following, however, I
shall point out another interesting effect relevant for this
epoch.

Together with the initial conditions $P_x(0) \simeq P_y(0) \simeq
0$, Eqs.\ (\ref{psolution}) and (\ref{angle}) give, in the low
temperature $|\beta|\gg D$ limit, the expression \cite{brick}
 \begin{equation}
 \label{parkelike}
P_z(t) \simeq \cos 2\theta_{m,D}(t) \cos 2 \theta_{m,D} (0)
F_{{\rm eff}}  P_z(0),
\end{equation}
where $F_{\rm eff} = \exp \int^t_0 k_3(t') dt'$ is generally
analytically insoluble.  However, supposing that the neutrino is
created and ``measured''  well before and after resonance (plus
other assumptions), one may write down an asymptotic solution:
\begin{equation}
\label{efficiency} F_{{\rm eff}} = \exp \left[- \frac{\pi \beta
D}{\left| \frac{d \lambda}{dt} \right|} \right]_{{\rm res}},
\end{equation}
so that Eq.\ (\ref{parkelike}) becomes, in probability language,
\begin{equation}
\label{prob} {\rm Prob}(\nu_{\alpha} \to \nu_s,\ t \gg t_{\rm
res}) \simeq \frac{1}{2} \left( 1 + F_{{\rm eff}} \right).
\end{equation}

Physically,  as the system crosses a resonance, collisions disrupt
the coherent flavour conversion process and  weaken the MSW
mechanism, the extent of which is parameterised by an {\it
efficiency factor} $F_{\rm eff}$. In a collisionless environment,
$F_{{\rm eff}}  =1 $,  and Eq.\ (\ref{prob}) gives ${\rm Prob}
\simeq 1$ as expected.

The factor $F_{\rm eff}$ compares the interaction length
$\ell_{\rm int} = 1/D$ with the physical resonance width
$\ell_{\rm res} = |\beta/\frac{d \lambda}{dt}|$. If $\ell_{\rm
res}
> \ell_{\rm int}$, maximal mixing persists for a sufficiently long
 time during which substantial collision-induced equilibration
may occur.
 An infinite  resonance width means $F_{{\rm
eff}} \to 0$ and the system emerges from the resonance with equal
$\nu_{\alpha}$ and $\nu_s$ distribution functions.

\section{Conclusion}

Large relic neutrino asymmetries can be generated from
active--sterile neutrino oscillations  for a wide range of
oscillation parameters.  These asymmetries may then  suppress
sterile neutrino production and modify primordial $^4{\rm He}$
synthesis. Approximate evolution equations obtained from the QKEs
offer much analytical insight on the nature of the asymmetry
growth.

\end{document}